\documentclass[preprint,superscriptaddress,showpacs]{revtex4-1}
\usepackage{graphicx}
\usepackage[thickqspace]{SIunits}
\newcommand{\tas}{$1T$-TaS$_2$}

\begin{document}

\title{Polaronic conductivity in the photoinduced phase of \tas}

\author{Nicky Dean}
\email[]{n.dean1@physics.ox.ac.uk}
\affiliation{Department of Physics, University of Oxford, Clarendon Laboratory, Parks Road, Oxford, OX1 3PU, UK}

\author{Jesse Petersen}
\affiliation{Department of Physics, University of Oxford, Clarendon Laboratory, Parks Road, Oxford, OX1 3PU, UK}
\affiliation{Max Planck Department for Structural Dynamics, University of Hamburg, Centre for Free Electron Laser Science, Notkestra\ss e 85, 22607 Hamburg, Germany}

\author{Daniele Fausti}
\affiliation{Department of Physics, University of Oxford, Clarendon Laboratory, Parks Road, Oxford, OX1 3PU, UK}
\affiliation{Max Planck Department for Structural Dynamics, University of Hamburg, Centre for Free Electron Laser Science, Notkestra\ss e 85, 22607 Hamburg, Germany}

\author{Ra'anan I. Tobey}
\affiliation{Department of Physics, University of Oxford, Clarendon Laboratory, Parks Road, Oxford, OX1 3PU, UK}
\affiliation{Condensed Matter Physics and Material Sciences Department, Brookhaven National Laboratory, Upton, New York, USA}

\author{Stefan Kaiser}
\affiliation{Max Planck Department for Structural Dynamics, University of Hamburg, Centre for Free Electron Laser Science, Notkestra\ss e 85, 22607 Hamburg, Germany}

\author{Lev Gasparov}
\affiliation{University of North Florida, Department of Chemistry and Physics, 4567 St. Johns Bluff Road, South Jacksonville, Florida 32224, USA}

\author{Helmuth Berger}
\affiliation{Institute of Physics of Complex Matter, \'Ecole Polytechnique F\'ed\'erale de Lausanne, CH-1015 Lausanne, Switzerland}

\author{Andrea Cavalleri}
\affiliation{Department of Physics, University of Oxford, Clarendon Laboratory, Parks Road, Oxford, OX1 3PU, UK}
\affiliation{Max Planck Department for Structural Dynamics, University of Hamburg, Centre for Free Electron Laser Science, Notkestra\ss e 85, 22607 Hamburg, Germany}

\date{\today}

\begin{abstract}
The transient optical conductivity of photoexcited \tas{} is determined over a three-order-of-magnitude frequency range. Prompt collapse and recovery of the Mott gap is observed. However, we find important differences between this transient metallic state and that seen across the thermally-driven insulator-metal transition. Suppressed low-frequency conductivity, Fano phonon lineshapes, and a mid-infrared absorption band point to polaronic transport. This is explained by noting that the photo-induced metallic state of \tas{} is one in which the Mott gap is melted but the lattice retains its low-temperature symmetry, a regime only accessible by photo-doping.
\end{abstract}


\maketitle

Low-dimensional dichalcogenides exhibit a wealth of competing phenomena, ranging from Mott and Peierls transitions \cite{Fazekas1980Charge,vanSmaalen2004Peierls} to spin and charge density waves \cite{Gruner1994Density,Clerc2007Fermi}, and superconductivity \cite{Sipos2008From}. Many such materials also display photoinduced phase transitions, wherein light pulses can create new transient states by driving ultrafast changes in the macroscopic properties \cite{Schmitt2008Transient,Perfetti2006Time}.

\tas{} is a quasi-two-dimensional transition metal dichalcogenide in which charge density wave (CDW) behaviour coexists with strong electron-electron correlations \cite{Wilson1975Chargedensity,Fazekas1979Electrical,Clerc2006Latticedistortionenhanced,Uchida1981Infrared,Gasparov2002Phonon}. At high temperatures \tas{} is metallic. Below 350 K, a periodic lattice distortion creates star-shaped clusters of tantalum atoms, causing the CDW to become nearly commensurate as domains of such clusters appear \cite{Sipos2008From,Clerc2006Latticedistortionenhanced}. Below $T_C = 180$ K, as the CDW becomes fully commensurate with the lattice and a long-range $\sqrt{13}\times\sqrt{13}$ superlattice structure emerges \cite{Fazekas1979Electrical,Clerc2006Latticedistortionenhanced,Uchida1981Infrared}, the system becomes insulating. This structural distortion reduces the bandwidth below a critical value, and leads to the opening of a 100-meV correlation (Mott) gap, with electrons localized at the star centres \cite{Clerc2007Fermi,Fazekas1979Electrical,Gasparov2002Phonon}. Mott localisation explains the low-temperature conductivity, which is lower than that expected from band structure \cite{Fazekas1979Electrical}.

The general features of the photoinduced dynamics in \tas were first revealed in all-optical studies \cite{Demsar2002Femtosecond,Toda2004Anomalous,Onozaki2007Coherent}, which highlighted important changes in the near-infrared optical properties and the excitation of coherent CDW amplitude mode oscillations. Time-resolved photoelectron spectroscopy experiments \cite{Perfetti2006Time} have provided a more comprehensive view of the physics, demonstrating the prompt collapse and recovery of the gap at $k = 0$ on a timescale short compared to the period of the coherent amplitude mode. This was interpreted as ultrafast melting of the Mott insulator, driven primarily by carrier temperature and only weakly influenced by the amplitude mode. Recent ultrafast electron diffraction measurements substantiate this view, showing that photo-doping does not trigger complete relaxation of the structural distortion \cite{Eichberger2010Ultrafast}.

Here, we set out to probe the electrodynamics of this transient Mott-melted phase, using a combination of ultrafast optical probes that cover more than three decades of frequency. In this way, we access a state of the solid in which photo-doping has removed the correlation gap, whilst the lattice symmetry, and presumably the bandwidth, of the low-temperature Mott insulator are only weakly perturbed.

In our experiments, \tas{} was held in its low-temperature insulating phase at 15 K and excited with 1.3-\micro\metre-wavelength pulses (950 meV) from an optical parametric amplifier. In a first set of measurements, low-energy transport was probed with time-resolved THz spectroscopy \cite{Nuss1991Dynamic,Beard2001Subpicosecond,Knoesel2001Charge}. The time-dependent reflectivity of the sample was referenced against that of the unperturbed sample and of a gold-coated portion of the surface. This allowed for the determination of the complex optical constants for each time delay \cite{Nashima2001Measurement}. The time-dependent conductivity was extracted by considering a thin photoexcited film with an unperturbed bulk volume beneath, thereby keeping into account the mismatch in pump and probe penetration depths (45 nm and 7 \micro\metre, respectively).

\begin{figure}[tb]
\centering
\includegraphics[width=\columnwidth]{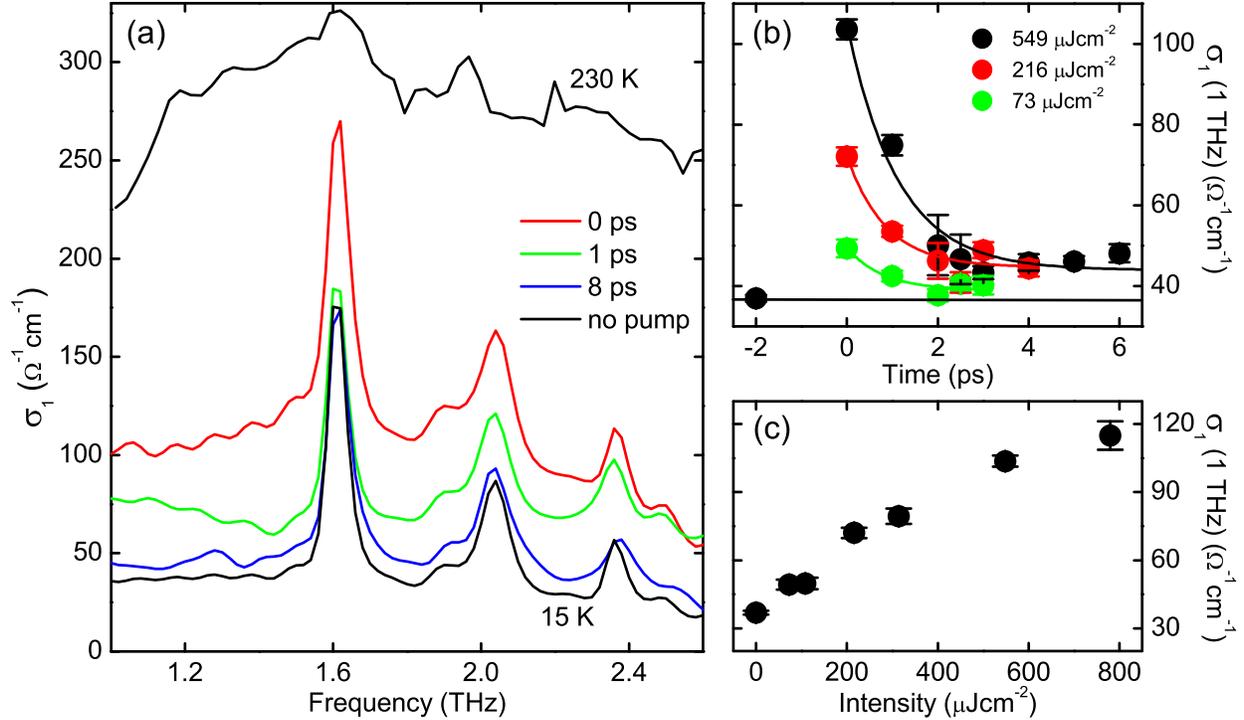}
\caption{(a) Real part of the conductivity $\sigma_1$ as a function of pump-probe delay for 550 \micro\joule{} \centi\rpsquare\metre{} pump intensity. Static conductivity in the low- and high-temperature phases is shown in black for comparison (high-temperature data from \cite{Gasparov2002Phonon}). (b) Transient behaviour of $\sigma_1$(1 THz) for different pump intensities (lines are guides to the eye). (c) Intensity dependence of $\sigma_1$(1 THz) at 0 ps time delay.}\label{fig:conductivity}
\end{figure}

Figure \ref{fig:conductivity}(a) shows the real part of the conductivity $\sigma_1$ obtained in the low-temperature equilibrium phase and at several pump-probe time delays. Drude-like behaviour is indicated by a flat response in $\sigma_1$. The temporal evolution of $\sigma_1$(1 THz) is plotted in Fig. \ref{fig:conductivity}(b). A prompt increase is followed by exponential decay back towards equilibrium, reflecting the loss and recovery of the Mott gap \cite{Perfetti2006Time}. However, the photoinduced change in $\sigma_1$ is significantly smaller than that observed upon heating. Furthermore, the three low-lying phonon modes \cite{Uchida1981Infrared,Gasparov2002Phonon} remain essentially unscreened, and $\sigma_1$ increases continuously with excitation fluence, as shown in Fig. \ref{fig:conductivity}(c). All these observations are inconsistent with the larger step-like increase in $\sigma_1$ and the complete screening of the phonon modes observed at 220 K \cite{Gasparov2002Phonon,Thompson1971Transitions}. This indicates a transient metallic state with qualitatively different transport properties to those observed in the thermal metallic state.

\begin{figure}[tb]
\centering
\includegraphics[width=\columnwidth]{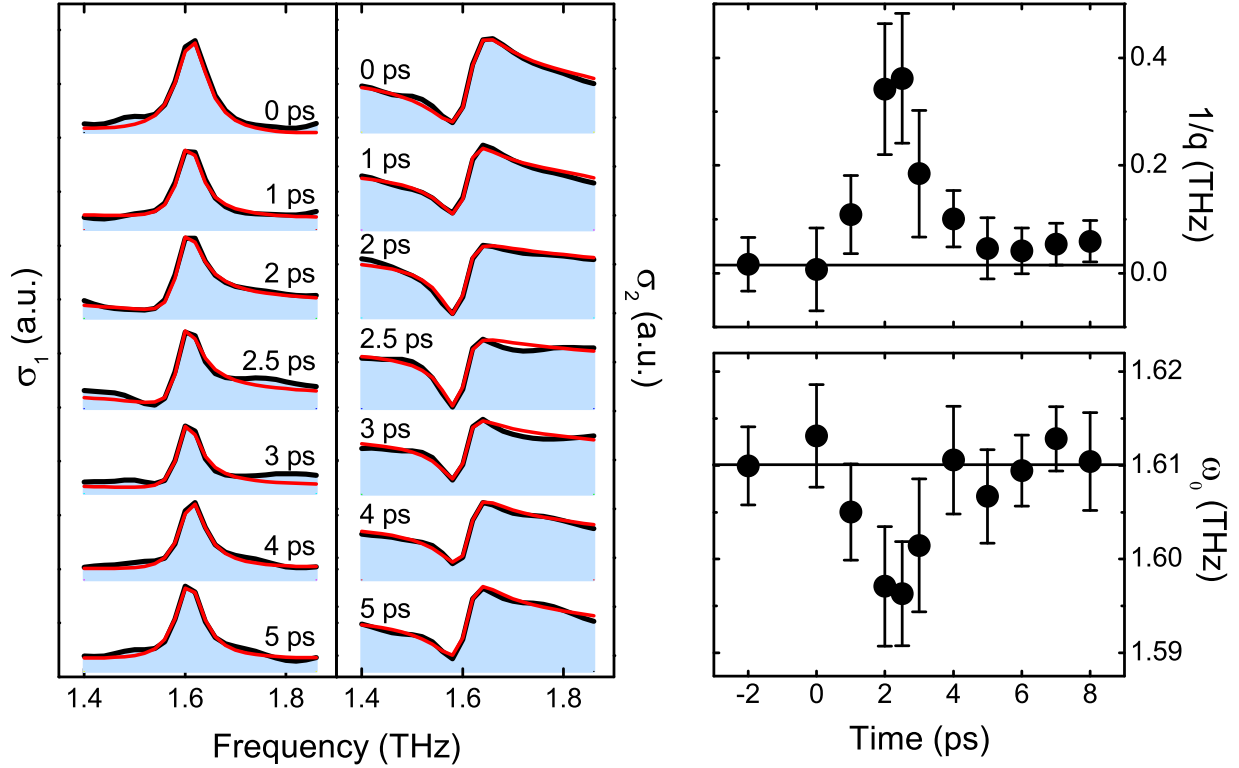}
\caption{Temporal evolution of the 1.6 THz mode after excitation with 550 \micro\joule{} \centi\rpsquare\metre{} pump pulse. Left panels: real and imaginary parts of the conductivity for different pump-probe delays (black lines). Red lines are fits using a Fano lineshape. Right panels: time-dependent asymmetry parameter $1/q$ and renormalized mode frequency $\omega$.}\label{fig:fano}
\end{figure}

Figure \ref{fig:fano} shows the complex conductivity around the 1.6 THz phonon resonance at various time delays. We find that the Lorentzian phonon lineshapes of the insulator become strongly asymmetric in the photoexcited state. The real part $\sigma_1$ exhibits a clear shift of spectral weight from the low- to the high-energy side of the mode, with a dip developing below the central frequency. The imaginary part $\sigma_2$ flattens on the high-energy side. A similar effect, though of smaller magnitude, is seen in the other two modes within our measurement bandwidth.

These asymmetries are well described by the Fano effect, a general spectroscopic feature which arises from coupling between a discrete resonance and a continuum of excitations \cite{Fano1961Effects}. Such coupling produces interference around the resonance, leading to a complex conductivity of the form
\[\sigma(\omega) = i\sigma_{0}(q-i)^2\left(i + \frac{\omega^2 - \omega_0^2}{\gamma_0\omega}\right)^{-1},\]
where $\omega_0$ and $\gamma_0$ are the mode central frequency and linewidth, respectively, and $q$ is the Fano asymmetry parameter ($1/q$ denotes a more asymmetric profile) \cite{Fano1961Effects,Kezsmarki2006Variation}. Fano effects in phonon lineshapes have been observed in other correlated electron systems both statically \cite{Kezsmarki2006Variation,Lupi1998Fano} and dynamically \cite{Hase2006Photoinduced,Lee2006Ultrafast}. Their presence indicates strong electron-phonon coupling, and studying them can reveal much useful information on the interactions in correlated systems \cite{Lupi1998Fano,Damascelli1997Infrared}.

The red lines in Fig. \ref{fig:fano} show a fit to the lineshape of the 1.6 THz mode as a function of pump probe time delay. The asymmetry parameter and renormalized phonon frequency, shown on the right-hand side of Fig. \ref{fig:fano}, exhibit the same temporal dependence as $\sigma_1$ in Fig. \ref{fig:conductivity} and of the Mott gap measured in Ref. \cite{Perfetti2006Time}. The sign of the asymmetry shows that the phonons are interacting with a continuum centred at higher energy \cite{Fano1961Effects,Lupi1998Fano}. Most likely this continuum is the gas of itinerant electrons.

\begin{figure}[tb]
\centering
\includegraphics[width=0.8\columnwidth]{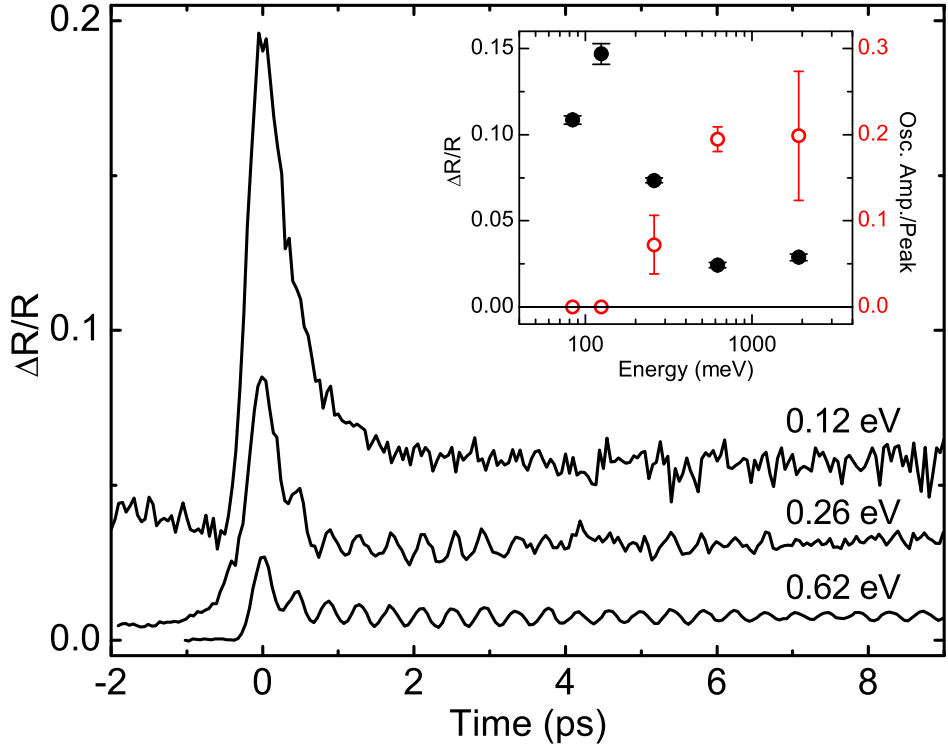}
\caption{Pump-probe traces for three probe photon energies. Inset: peak reflectivity changes ($\Delta R/R$) as a function of probe photon energy (black dots, left axis) and ratio of oscillation amplitude to peak $\Delta R/R$ (red circles, right axis). In all cases the pump intensity was $\sim$ 750 \micro\joule{} \centi\rpsquare\metre{}}\label{fig:pumpprobe}
\end{figure}

At higher photon energies, photoinduced reflectance changes were measured with pulses from a second optical parametric amplifier, continuously tuneable from the 100-meV Mott gap (12.4-\micro\metre wavelength) to the visible. Figure \ref{fig:pumpprobe} shows typical time-domain traces obtained for three different probe energies above the gap. The reflectance shows a sudden increase near zero time delay, followed by decay with superimposed 2.4-THz oscillations from the coherent CDW amplitude mode \cite{Demsar2002Femtosecond,Toda2004Anomalous,Onozaki2007Coherent}.

The magnitude of the reflectance change, the relative size of the oscillations, and their lifetime, all depend on the probe photon energy, as illustrated in the inset of Fig. \ref{fig:pumpprobe}. As the probe energy is lowered towards the gap, $\Delta R/R$ increases dramatically due to the shift of spectral weight into the previously gapped region \cite{Perfetti2006Time}. Conversely, the coherent lattice vibrations influence the optical response progressively less. This suggests that the structural distortions are only weakly connected to the low-frequency electrodynamics, further substantiating a view in which the Mott gap is melted independently from the coherent CDW distortions.

\begin{figure}[b]
\centering
\includegraphics[width=0.9\columnwidth]{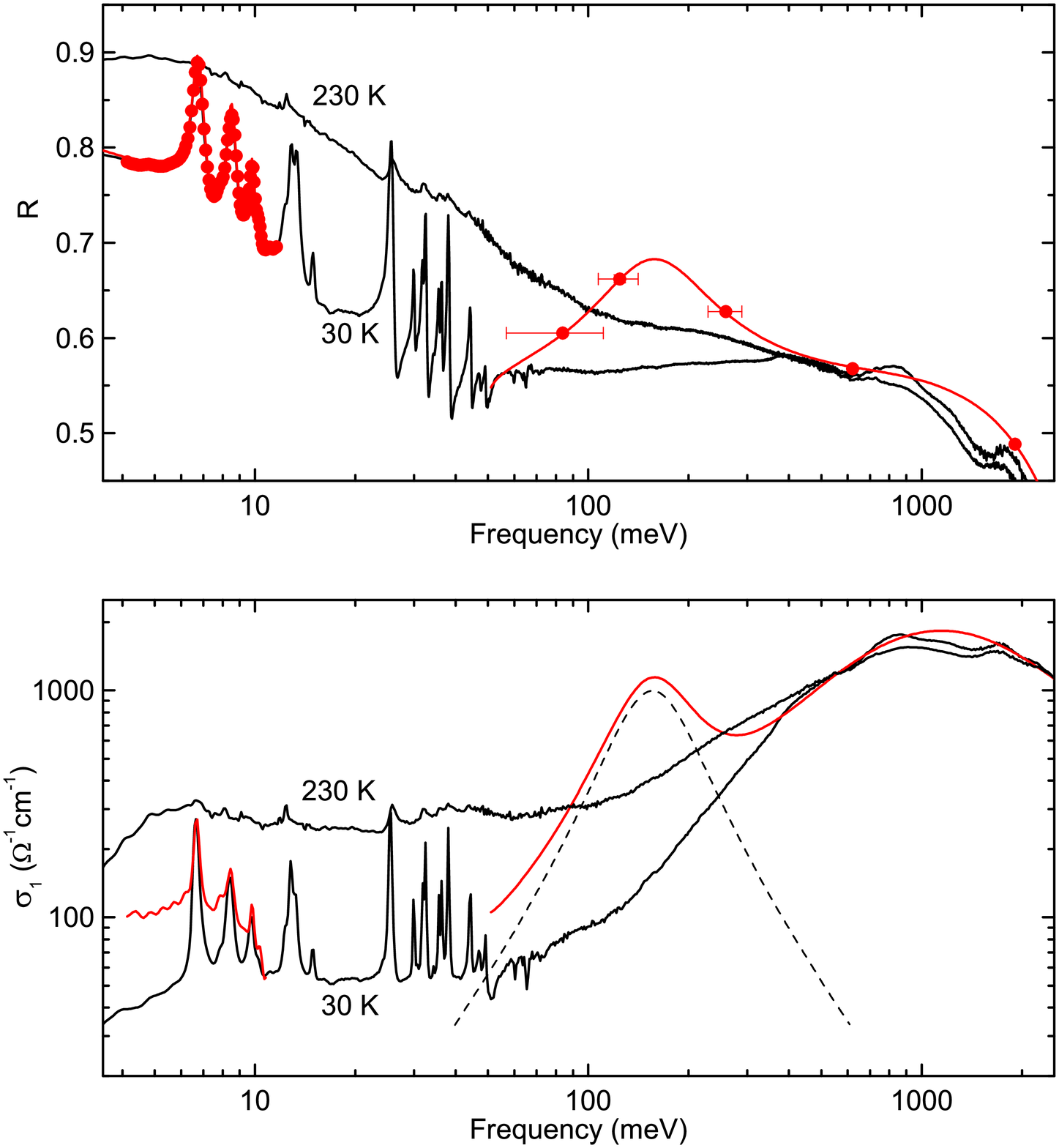}
\caption{Upper panel: reflectance change immediately after photoexcitation (red dots). Red line is a fit to the data (see text). Lower panel: calculated conductivity immediately after photoexcitation (red line). Dashed line is a Lorentzian representing a polaron band. In both panels, continuous black lines are the reflectance and conductivity of the thermal phases \cite{Gasparov2002Phonon}.}\label{fig:fullfreq}
\end{figure}

The photoinduced changes in reflectance from the THz to the visible are summarised in Fig. \ref{fig:fullfreq}. Note that whilst in the THz range the raw signal is small ($\ll1\%$) due to the large difference between the pump and probe penetration depths, in the mid-infrared region, where the extinction coefficients are better matched, a pronounced increase in reflectance is clearly visible already in the unprocessed data.

To extract broadband time-dependent optical conductivity in absence of phase information at high photon energies, we applied the following procedure. First, the static broadband reflectance at 30 K was fitted by modelling the equilibrium optical conductivity with a series of Lorentzians, accounting for the phonons and electronic transition bands. Secondly, we analysed the photoinduced changes in reflectivity, which could be modelled by shifts of spectral weight among the oscillators of the equilibrium fit. However, the mid-infrared response required the addition of a new resonance centred at 155 meV. This feature has no analogue in any of the thermal phases.

The lower panel of Fig. \ref{fig:fullfreq} shows the broadband conductivity of the photoinduced phase at 0 ps time delay, as determined by this procedure. Three features emerge as especially meaningful: reduced low-frequency conductivity compared to the high-temperature phase, Fano lineshapes for the phonons, indicative of strong electron-lattice coupling, and a transient mid-infrared resonance. The nature of this mid-infrared band is consistent with that of polaron bands in other strongly correlated materials \cite{Lupi1999Evolution,Ciuchi2008Signatures}, and the asymmetric reshaping in the phonon lineshapes emphasises strong coupling between electrons and the lattice coinciding with the appearance of this new feature.

From the combined observations above, then, the key conclusion of our work is that transport in the photoinduced phase is polaronic. Microscopically, charges that in the ground state were localized by electron correlations are made mobile after photo-doping, although electron-lattice coupling, which is still strong due to the presence of the CDW, imposes a dominant energy scale on their transport. These features belong to an exotic state of \tas{} that is only accessible with photo-doping.

In summary, we have used optical spectroscopy over a broad frequency range to measure the transport properties of photoexcited \tas. Our data clearly indicate the formation of a new state, with polaronic transport that descends directly from the separate influence that light has on electron correlations and on the lattice response. As such, the use of ultrabroadband time-resolved optical spectroscopy emerges as an indispensable tool to understand the role of many interacting degrees of freedom in transient states of complex matter. This dissection of the relative importance of electron and lattice contributions may point towards a better understanding of the formation of the Mott-CDW ground state in this intriguing material.

Correspondence should be addressed to Nicky Dean (n.dean1@physics.ox.ac.uk) and to Andrea Cavalleri (andrea.cavalleri@mpsd.cfel.de).
\bibliography{paper}

\end{document}